# Algorithm for Finding an Exact Maximum Distance in $E^2$ with $O_{exp}(N)$ Complexity: Analysis and Experimental Results

VACLAV SKALA [0000-0001-8886-4281]

Dept. of Computer Science and Engineering, Faculty of Applied Sciences, University of West Bohemia,
CZ 301 00 Pilsen, Czech Republic

Finding a maximum distance of points in $E^2$ or in $E^3$ is one of those. It is a frequent task required in many applications. In spite of the fact that it is an extremely simple task, the known "Brute force" algorithm is of $O(N^2)$ complexity. Due to this complexity the run-time is very long and unacceptable especially if medium or larger data sets are to be processed. An alternative approach is convex hull computation with complexity higher than $O(N)$ followed by diameter computation with $O(M^2)$ complexity. The situation is similar to sorting, where the bubble sort algorithm has $O(N^2)$ complexity that cannot be used in practice even for medium data sets.

This paper describes a novel and fast, simple and robust algorithm with $O(N)$ expected complexity which enables to decrease run-time needed to find the maximum distance of two points in $E^2$. It can be easily modified for the $E^k$ case in general. The proposed algorithm has been evaluated experimentally on larger different datasets in order to verify it and prove expected properties of it.

Experiments proved the advantages of the proposed algorithm over the standard algorithms based on the "Brute force", convex hull or convex hull diameters approaches. The proposed algorithm gives a significant speed-up to applications, when medium and large data sets are processed. It is over 10 000 times faster than the standard "Brute force" algorithm for $10^6$ points randomly distributed points in $E^2$ and over 4 times faster than convex hull diameter computation. The speed-up of the proposed algorithm grows with the number of points processed.

Categories and Subject Descriptors: I**.3.5.[Computer Graphics]:** Computational Geometry and Object Modeling – geometric algorithms, languages, and systems

## 1. INTRODUCTION

A maximum distance of two points in the given data set is needed in many applications. A standard "Brute Force" algorithm with $O(N^2)$ complexity is usually used, where $N$ is a number of points in the given data set. Such algorithm leads to very high run-time if larger data sets are to be processed. As the computer memory capacity increases, larger data sets are to be processed. Typical data sets in computer graphics contain usually $10^5$-$10^7$ and even more of points. In spite of the CPU speed increases, the run-time even for such a simple task leads to unacceptable processing time for today's applications. Of course, there is a very special case when points are distributed on a circle only. This requires the $O(N^2)$ algorithm if we want to find **all the couples** of points as there is $N(N-1)/2$ couples. In all other cases "output sensitive" algorithms should be faster.

Nevertheless our task is just to find the maximum distance, not all the pairs having a maximum distance. So the complexity of this algorithm should be lower. Also due to the numerical precision points do not lie exactly on a circle if data have this very specific property.

The new proposed algorithm with $O(N)$ expected complexity is based on the following assumptions:
- Any pre-processing with a lower complexity than the optimal run-time one should speed-up processing of the given data set. In our case the optimal algorithm covering all the special cases is of $O(N^2)$ complexity and therefore preprocessing with complexities $O(lgN)$, $O(N)$, $O(N\ lgN)$ etc. should speed up the run-time.
- General properties, including geometrical ones, of input data should be carefully analyzed in order to find all useful information that can lead to faster pre-processing and the final run-time.
- If data are not organized in a very special way, e.g. points are on the Axis Aligned Bounding Box (AABB) boundary only or points are on a circle etc., we can use an algorithm with "output sensitive" complexity and we should get additional speed-up.

In general, algorithms should not depend on very specific presumptions or technological issues unless the algorithm is targeted to very specific technological platform or applications. Any algorithm must be stable and robust to input data properties, in general.

## 2. BRUTE FORCE ALGORITHM

The standard "Brute Force" algorithm uses two nested loops in order to find a maximum distance. Algorithms with such approach can be found in many text-books dealing with fundamental algorithms and data structures, e.g. Hilyard and Theilet [2007], Mehta and Sahni [2005], Sahni [1998], Sedgwick [2002], Wirth [1976]. Such algorithms can be represented by Alg.1 in general as:

```
function distance_2 (A , B: point);
{  distance_2:=(A.x-B.x)*(A.x-B.x) + (A.y-B.y)*(A.y-B.y)};
# Square of the distance ‖ A - B ‖ is actually computed #

d := 0;
for i := 1 to N-1 do
 for j := i+1 to N do
  {
    d0 := distance_2(Xi , Xj);
    if d < d0 then d := d0
  };
d := SQRT (d)  # if needed #
```
<center>Standard "Brute Force" algorithm<br>Algorithm 1</center>

The Alg.1 is clearly of $O(N^2)$ complexity and processing time increases significantly with number of points processed, see Tab.1.

In practice, it can be expected that points are not organized in a very specific manner, e.g. points on a circle etc., and points are uniformly distributed more or less. In this case "output sensitive" algorithms usually lead to efficient solutions.

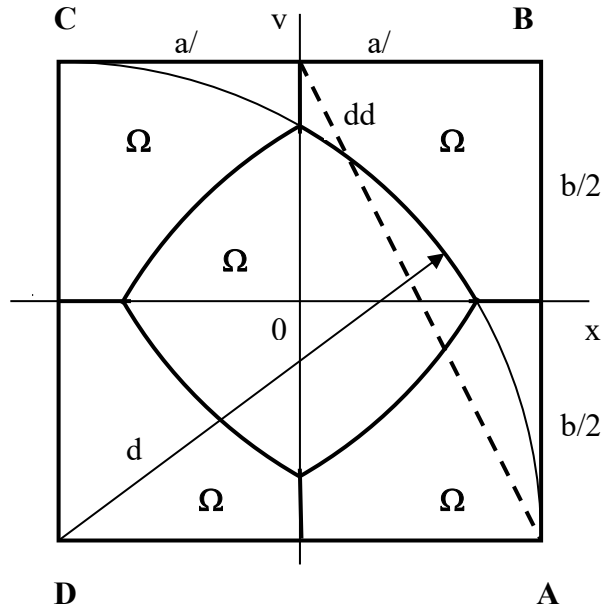

Fig.1. Splitting the $\Omega_0$ set to $\Omega_i$ sets for the worst case – squared area

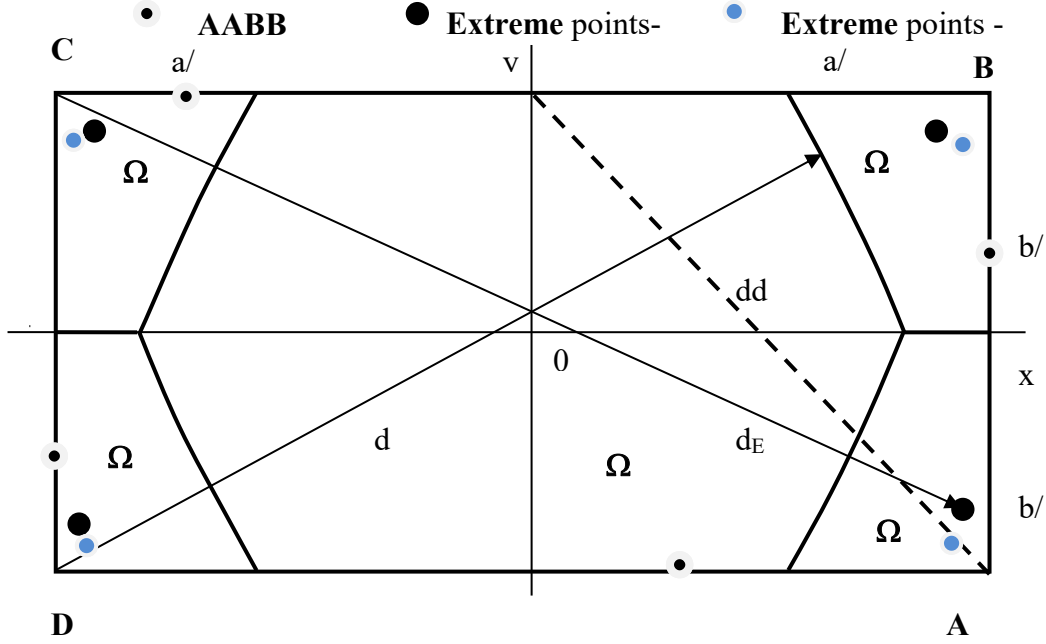

Fig. 2. Splitting the $\Omega_0$ set to $\Omega_i$ sets for the rectangular area case

Let points are inside of an Axis Aligned Bounding Box (AABB) defined as $<-a/2, a/2> \times <-b/2, b/2>$. Then Fig.1 presents a typical situation for the worst case when AABB is a square ($a = b$), while the Fig.2 presents general AABB situation for the case $a > b$. In the following we will explore the worst case, i.e. situation at the Fig.1, and the first maximum distance estimation $d$ is $d = a$.

It can be seen that points in the set $\Omega_0$ cannot influence the maximum distance computation in the given data set. We can remove all points $\Omega_0$ from the given data set $\Omega$ and obtain faster algorithm. As the maximum distance finding algorithm is of $O(N^2)$ complexity an algorithm with a lower complexity can be used in order to find and eliminate points which cannot influence the final distance. Space subdivision techniques can be used to split points into the disjunctive data sets $\Omega_i$ and decrease run-time complexity again. For a general case, when AABB is not squared, the $\Omega_0$ set will contain more points of course, see Fig.3.

Let us explore the worst case more in a detail, now.

$$d_1 = b^2 + \left(\frac{a}{2}\right)^2 \qquad d_2 = b^2 + a^2$$

Let us consider the case, when $a = k\,b$. Then $\xi^2 = d^2 - b^2$, $L = a - \xi$ and $L^2 = a^2 + \xi^2 - 2a\xi$.

Area $P$ is given as

$$P = \frac{1}{2}L^2 = \frac{1}{2}\left[a^2 + d^2 - b^2 - 2a\sqrt{d^2 - b^2}\right] = \frac{1}{2}\left[k^2 b^2 + d^2 - b^2 - 2kb\sqrt{d^2 - b}\right] =$$

$$\frac{1}{2}b^2\left[k^2 - 1 + \frac{d^2}{b^2} - 2k\sqrt{\frac{d^2}{b^2} - 1}\right]$$

If the most consuming parts with $O(N^2)$ complexity is considered, then the speed-up of the proposed algorithm over the "Brute Force" algorithm for uniformly distributed points is defined as:

$$\nu = \left[\frac{ab}{4P}\right]^2 = \frac{k^2 b^2 b^2}{b^2 b^2} \frac{k^2}{4\left[k^2 - 1 + d^2 - 2k\sqrt{d^2 - 1}\right]^2} = \frac{k^2}{4\left[k^2 - 1 + d^2 - 2k\sqrt{d^2 - 1}\right]^2} \quad \text{for } d_1 \le d \le d_2$$

For $k = 1$ and $d \to \sqrt{2}$ the speed-up $\nu \to \infty$ that is expected from the algorithm specification.

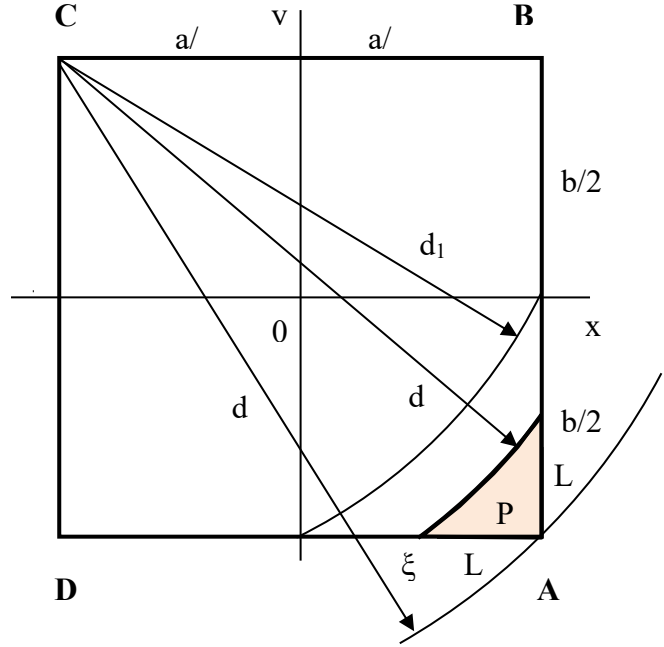

Fig. 3. Distances definitions for a general AABB

It can be seen that for a non-squared AABB distances are defined as

Table 1

| | | |
|---|---|---|
| $E^2$ | $d_1^2 = [\max\{a,b\}]^2 + \frac{1}{2}[\min\{a,b\}]^2$ | $d_2 = a^2 + b^2$ |
| $E^3$ | $d_1^2 = [\max\{a,b,c\}]^2 + \frac{1}{2}[\min\{a,b,c\}]^2$ | $d_2 = a^2 + b^2 + c^2$ |

$a$, $b$ are sizes of the AABB in $E^2$, resp. $a$, $b$, $c$ are sizes of the AABB in $E^3$

For the $E^3$ case, some minor changes have to be made as we have 6 points defining the AABB and 6+6 extreme points in the AABB, i.e. points having the longest distance and the shortest distance from the relevant AABB corner and the $\Omega_0$ set is to be split is to be split to sets $\Omega_0,..., \Omega_6$. However the computational time is more or less the same as the same number of points is processed and the preprocessing is of *O(N)* complexity. It can be seen that the extension to $E^k$ is straightforward and simple to implement.

**3. CONVEX HULL DIAMETER**
The idea of finding maximum distance of points by more effective algorithms is not new. Reasonably effective approach is based on a convex hull construction of the given data set. Then the convex hull points are processed by the standard algorithm with *O(N²)* complexity in order to find the maximum distance. It is obvious that that all techniques based on the convex hull construction have the following properties:
- Convex hull construction algorithms for a higher dimension than $E^2$, i.e. for $E^3$ or $E^k$ in general, are complex and quite difficult to implement. Skiena [1997] proved that the "gift wrapping algorithm" has $O(n^{k/2+1})$ complexity in the case of *k* dimensional problem. Yao [1981] has proved that for the two dimensional case specialized algorithm has *O(N lgN)* complexity.
- Points of the convex hull are to be processed by the **final** algorithm with *O(h²)* complexity, where *h* is the number of points of the computed convex hull, i.e. the technique is an output sensitive. The number of convex hull points might be quite high, while the maximum distance is usually given by two points in the given data set.

Generally, the well known algorithms have the computational complexities as follow: Brute Force *O(N⁴)*, Gift Wrapping *O(N h)*, Graham Scan *O(N lgN)*, Jarvis March *O(N lgN)*, Quick Hull *O(h N)*, Divide-and-Conquer *O(N lgN)*, Monotone Chain *O(N lgN)*, Incremental *O(N lgN)*, Marriage-before-Conquest *O(n lgh),* see Barber [1996], O'Rourke [1998], Yao [1981], Kirkpatrick [1986], Chan [1996], Avis [1997], WEB [1].

Some algorithms directed to the diameter of a convex hull computation can be found in Snyder [1980], Dobkin [1979], Shamos [1978]. It should be noted that if the number $h$ of the resulted convex hull is close to $N$, than algorithms with the complexity $O(h N)$ are becoming algorithms with $O(N^2)$ complexity etc.

The extension to a higher dimension is not easy and some algorithms cannot be extended even for $E^3$, e.g. Graham Scan etc. or the complexity of the actual implementation is prohibitive for practical use.

## 4. PROPOSED ALGORITHM

The new proposed algorithm was developed for larger data sets and it is based on "in-core" technique, i.e. all data are stored in a computer memory. The fundamental requirements for the algorithm development were: simplicity, robustness and simple extensibility to $E^3$. The proposed algorithm is based on two main principles:
- Remove as many non-relevant points as possible
- Divide and conqueror technique in order to decrease algorithm complexity

Fig.1 shows five regions $\Omega_i$, where the given points are located. The algorithm is described for the $E^2$ case and its extension to $E^3$ is straightforward. It should be noted that the worst case is presented, i.e. when AABB is a square. Let us assume that the points that cannot contribute to the final maximum distance are located in the region $\Omega_0$, which contains points closer to all corners of the AABB than the *minimal* edge length of the AABB or known distance estimation. Then the given data set $\Omega$ can be reduced to $\Omega = \Omega - \Omega_0$. In order to decrease expected number of points to be processed, we need to process this data set $\Omega$ to get more information on those points. As the standard algorithm for maximum distance is of $O(N^2)$ complexity, we can use any pre-processing of $O(N)$ or $O(N \lg N)$ complexity to decrease number of points to be left for the final processing with the algorithm of $O(N^2)$ complexity.

It can be seen that the following principal steps have to be made:
1. Pre-processing: can be performed with $O(N)$ complexity:
    a. Find the bounding AABB and extreme points, i.e. two extreme points for each axis (max. 4 points).
    b. Find the most distant "extreme" points [max] for each corner of the AABB (max. 4 points).
    c. Find the minimum distant "extreme" points [min] for each corner of the AABB (max. 4 points).
    d. Determine the longest mutual distance $d$ between those found points (max.12 points).
        It should be noted that the worst case is a squared AABB and found distance $d \geq a$.
        For a rectangular window found distance $d \geq max\{a,b\}$.
    e. Determine points of $\Omega_0$ that cannot contribute to the maximum distance, i.e. points having a smaller distance than the found distance $d$ from all corners of the AABB and extreme points. Remove the $\Omega_0$ points from the original data set $\Omega$.
    f. Split remaining points to new sets $\Omega_i$, i=1,..,4, see the Fig.1.

The number $S_{\Omega 0}$ of points in the $\Omega_0$ set can be estimated as:
$$S_{\Omega 0} = \int_{a/2}^{a\sqrt{3}/2} \sqrt{a^2 - x^2}\, dx = a^2 (\frac{\pi}{3} - \sqrt{3} + 1)$$

For the uniform distribution of points the $\Omega$ set, the number of points to be processed, i.e. number of points outside of the $\Omega_0$ set, is $qN = 0,684\ N$, where:
$$q = \frac{S_\Omega - S_{\Omega 0}}{S_\Omega} = \frac{a^2 - a^2(\frac{\pi}{3} - \sqrt{3} + 1)}{a^2} = \frac{\pi}{3} - \sqrt{3} \cong 0.684$$

As the "Brute Force" algorithm is of the $O(N^2)$, the speed up expected is approx.:
$$v = 1/(0.684)^2 = 2.13$$

It should be noted that the distance $d >> a$ in practical data sets and the $\Omega_0$ set contains much more points which can be removed from the final processing, see Tab.1 actually compared points, i.e. last column.

It can be seen that if found distance $d \geq dd$, see Fig.1, then the comparison of points from the neighbor data sets is not needed, where:
$$dd = \sqrt{\left(a^2 + (b/2)^2\right)} \leq \sqrt{a^2 + a^2/4} = \frac{\sqrt{5}}{2} a \quad \text{for} \quad a \leq b$$

2. Run-time steps of the proposed algorithm:
    a. Taking an advantage of space subdivision, find the maximum distance $d$ between points of $[\Omega_1, \Omega_3]$, i.e. one point from $\Omega_1$ and the second point is from $\Omega_3$ as there can be expected the longest distance between the given points - this step is $O(N^2)$ complexity.
    b. Remove points from the $\Omega_2$ and $\Omega_4$ datasets closer to the related corner of the AABB than already found distance $d$ - this step is $O(N)$ complexity.
    c. Find a new maximum distance $d$ between points of $[\Omega_2, \Omega_4]$ - this step is $O(N^2)$ complexity.
    d. If already found distance $d \leq dd$ then
        i. Reduce $\Omega_1, \Omega_2, \Omega_3, \Omega_4$ - steps are $O(N^2)$ complexity
        ii. find a new maximum distance $d$ between points of $[\Omega_1, \Omega_2], [\Omega_2, \Omega_3], [\Omega_3, \Omega_4]$ and $[\Omega_4, \Omega_1]$. It is necessary to note that if $d > dd$, then the $\Omega_0$ boundary crosses the AABB and points in the neighbors regions cannot contribute to the maximum distance.

As can be seen the algorithm is very simple and easy to implement.

```
function distance_2 (A , B: point);
{  distance_2:=(A.x-B.x)*(A.x-B.x) + (A.y-B.y)*(A.y-B.y)};
# Square of the distance ‖A - B‖ is actually computed #

function S_Dist (ΩA , ΩB : set)
{  d := 0; d0 := 0;
   for each point X from ΩA do
      for each point Y from ΩB do
      {  d0 := distance_2 (X , Y);   if d0 > d then d := d0
      };
      S_Dist := d
}
```

1. **Q** := points forming the AABB for the given set $\Omega$ and extreme points [max and min] **XX** for each corner of the AABB.
   # 8 points found at maximum, complexity $O(N)$ #
2. $d_M$=max { $Q_i$ , $Q_j$ }
   # Determine the maximum distance $d_M$ of the points in **Q** #
   # by the "Brute Force" algorithm with $O(M^2)$ complexity#
   # only max. 8 points are to be processed #
3. # the set $\Omega$ is to be split into $\Omega_i$ sets#
   **for all** points **X** from the set $\Omega$
   {  i := index of the region $\Omega_i$ for the point **X**
      $d$ = distance of the point **X** and of the opposite AABB corner for the set $\Omega_i$.
      # do not store points having higher distance from a AABB corner than $d_M$ #
      **if** $d \geq d_M$ **then**
      { STORE ( **X** , $\Omega_i$ ); # store a point **X** in the $\Omega_i$ set #
        $d_M := d$ ; # update the maximum distance $d_i$ for the region i#
        $XX_i := X$ # update $XX_i$ – one extreme point for each region $\Omega_i$ #
      }
   }
4. # new maximum distance estimation based on extreme points of sets $\Omega_i$ found in step 3#
   $d_q$ = max { $XX_i$ , $XX_j$ },   i, j =1,…,4
   $d_M$ = max { $d_M$ , $d_q$ }

5. # The "diagonal" regions are to be tested with $O(N^2)$ algorithm #
   # REDUCE ( $\Omega_i$ , $d_M$ ) remove points from the $\Omega_i$ set with smaller distance from the opposite AABB corner #
   REDUCE ( $\Omega_1$ , $d_M$ );      REDUCE ( $\Omega_3$ , $d_M$ );
   $d_M$ := max { $d_M$ , S_Dist( $\Omega_1$ , $\Omega_3$ ) };
   REDUCE ( $\Omega_2$ , $d_M$ );      REDUCE ( $\Omega_4$ , $d_M$ );
   $d_M$ := max { $d_M$ , S_Dist( $\Omega_2$ , $\Omega_4$ ) }

6. # neighbor regions should be tested if necessary #
   **if** $d_M \leq dd$ **then**
   { REDUCE ( $\Omega_1$ , $d_M$ ); REDUCE ( $\Omega_2$ , $d_M$ );    REDUCE ( $\Omega_3$ , $d_M$ ); REDUCE ( $\Omega_4$ , $d_M$ );
   $d_M$ := max { $d_M$ , S_Dist( $\Omega_1$ , $\Omega_2$ ) };    $d_M$ := max { $d_M$ , S_Dist( $\Omega_2$ , $\Omega_3$ ) };
   $d_M$ := max { $d_M$ , S_Dist( $\Omega_3$ , $\Omega_4$ ) };    $d_M$ := max { $d_M$ , S_Dist( $\Omega_4$ , $\Omega_1$ ) }
   }
7. $d$ := SQRT ($d_M$)    # compute the final distance as a square root of $d$ #

Algorithm 2. Fast maximum distance algorithm

**Implementation notes**

There are several possibilities how to further improve the proposed algorithm especially in the context of the specific programming language and data structures used. Nevertheless, the influence of this is small as experiments proved and for the expected data sizes do not have any significant influence.

Generally, it is recommended:

- The "array list" construction should be used for storing $\Omega_i$ sets; this construction enables to increase an array size without reallocation and data copying,
- Two or higher dimensional arrays for storing x, y values should not be used, as for each array element one addition and one multiplication operations are needed (computational cost is hidden in the index evaluation). Data should be stored in two arrays **X** and **Y**, or as pairs (x, y) in one-dimensional structure array **XY** etc.
- Square of a distance should be used in order to save multiple square root evaluations. It is possible as the square function is monotonically growing and can be used for comparison operations.
- Store data in a linked list as only a sequential pass is required and remove, resp. insert operation is simple and no data overwriting is required.

It should be noted that the $\Omega_i$ sets are determined by an arc of a circle, i.e. the separation function is quadratic. Experiments proved that if a half-space separation function is used, the proposed algorithm is faster as only a linear function is evaluated.

## 5. EXPERIMENTAL RESULTS

The standard "Brute Force", convex hull (Quick Hull) and proposed algorithms were implemented in C# and Pascal/Delphi, verified and extensively tested for different sizes of the given data sets and different data set types (random, uniform, clustered etc.) as well. Standard PC with 2,8 GHz Intel Pentium 4, 1 GB RAM with MS Windows XP was used.

Cumulative results obtained during experiments are presented in Tab.2. Experiments made proved that a significant speed-up has been reached. It is necessary to note that the speed-up 100 means that the computation is *100 times faster*. It can be seen that for $10^6$ points the speed-up is 10 000, i.e. computation is **$10^4$ times faster** and grows with the number of points nearly exponentially, see Fig.5 - note that axes scale is logarithmic.

Table 2 Experimental results for uniformly distributed points [* values obtained by extrapolation]

| Points $10^3$*N | Computational time [ms] | | | Speed-up | | | Compared | |
|---|---|---|---|---|---|---|---|---|
| | Brute Force (BF) | Quick Hull (QH) | New | BF/QH | BF/New | QH/New | QH | New |
| 100 | 137 760 | 108 | 15 | 1 281 | 9 462 | 7,38 | 405,1 | 16,1 |
| 160 | 353 920 | 178 | 25 | 1 987 | 14 364 | 7,23 | 454,0 | 23,7 |
| 250 | 865 760 | 260 | 56 | 3 332 | 15 460 | 4,64 | 486,0 | 26,9 |
| 400 | 2 216 480 | 451 | 84 | 4 911 | 26 387 | 5,37 | 583,1 | 26,1 |
| 630 | 5 498 080 | 720 | 167 | 7 635 | 32 946 | 4,32 | 661,2 | 32,7 |
| 1 000 | 13 783 840 | 1 130 | 259 | 12 197 | 53 277 | 4,37 | 707,1 | 25,3 |
| 1 600 | 35 467 040 | 1 721 | 364 | 20 603 | 97 437 | 4,73 | 735,9 | 30,8 |
| 2 500 | 86 591 680 | 3 069 | 664 | 28 217 | 130 378 | 4,62 | 761,9 | 22,3 |
| 4 000 | 221 673 760 * | 5 523 | 1 231 | 40 139 | 180 094 | 4,49 | 751,2 | 23,1 |
| 6 300 | 507 556 000 * | 9 544 | 1 708 | 53 183 | 297 164 | 5,59 | 750,0 | 23,5 |
| 10 000 | 1 106 173 600 * | 15 064 | 2 470 | 73 432 | 447 916 | 6,10 | 750,0 | 23,2 |

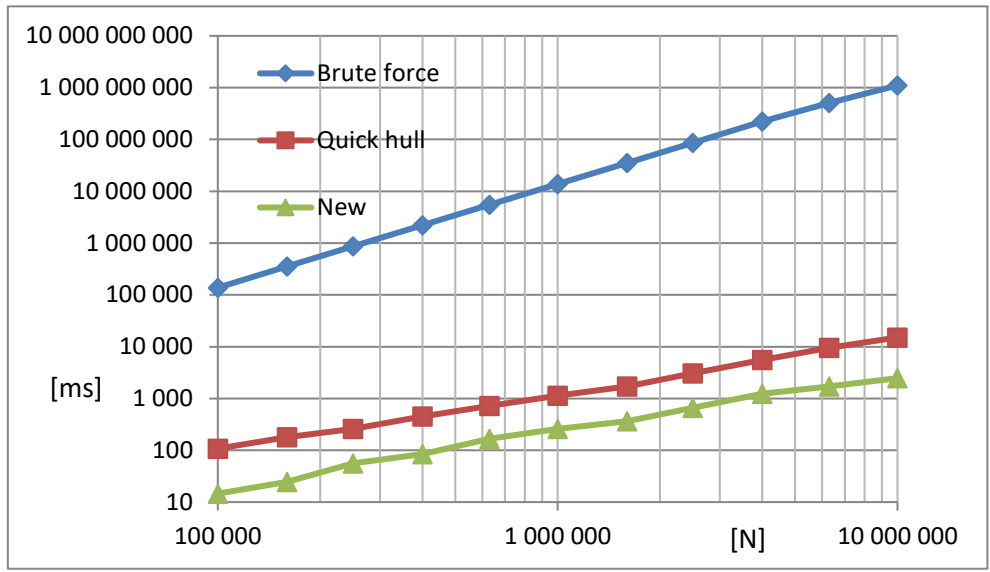

Fig. 4. Computational time of the "Brute force", Quick hull and the proposed algorithm

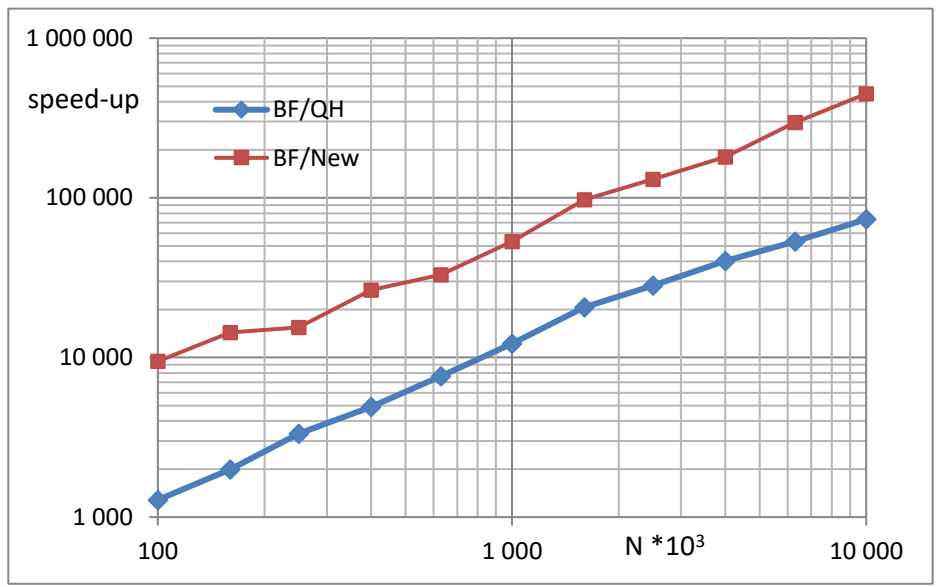

Fig. 5. Speed up of the Quick hull and the proposed algorithm over the Brute force algorithm

The experiments proved that the speed-up grows significantly with the number of points processed, see Fig.4 and Fig.5. The final step of the proposed algorithm of $O(N^2)$ complexity has a low influence and that the preprocessing steps significantly decrease number of points processed in the final step. This is due to the very low number of points remaining for the final evaluation for maximum distance, see Tab.1, where "QH" presents number of points finally processed after construction by the Quick Hull method, while "New" presents number of points finally processed by the proposed algorithm.

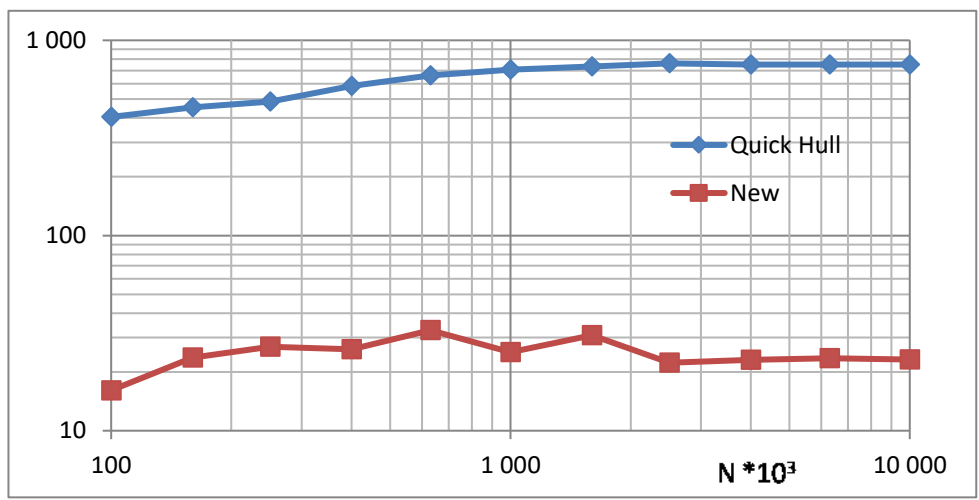

Fig 6 Number of points remaining for the final processing by the "Brute Force" algorithm

Several convex hulls algorithms were used in order to compare efficiency of the proposed algorithm. The convex hull based algorithms in $E^2$ proved reasonable results but the proposed algorithm was at least 4-5 times faster than algorithms based on the convex hull approach. The proposed algorithm was originally intended for $E^2$ and $E^3$ applications, but it is easily extendible for the $E^k$ case as well. Experiments proved robustness and faster computation of the proposed algorithm for data sets with different characteristics, i.e. Gaussian distribution, clusters and etc.

The speed-up over the convex-hull approaches is primarily caused by "ordering" data into $\Omega_i$ sets made with $O(N)$ complexity and also all other steps are of $O(N)$ complexity. Only the final computation is of $O(N^2)$ complexity, but the number of data processed by the proposed algorithm is rather small, see Tab.1. On the opposite the convex hull construction has a higher complexity than $O(N)$ in general and number of points left for the final processing with $O(N^2)$ complexity has approx. 10 times more points left for processing, i.e. 100 times more computations actually due to the final step with $O(N^2)$ complexity.

## 6. CONCLUSIONS

A new simple, easy to implement, robust and effective algorithm for finding a maximum distance of points in $E^2$ was developed. The experimental results clearly proved that the proposed algorithm is convenient for medium and large data sets. Algorithm speed-up grows significantly with the number of points processed. The proposed algorithm can be easily extended to $E^3$ by a simple modification. In the $E^3$ case, we have to process $\Omega_1, \ldots, \Omega_8$ data subsets. For the $E^k$ case the original data must be split to data sets $\Omega_i$, where $i = 1,\ldots,2^k$. Nevertheless the memory requirements and preprocessing time remain the same as we have to only split data from the $\Omega$ set to the to $\Omega_i$ datasets which are smaller.

The experiments also proved that the algorithm offers higher speed-up than algorithms based on convex-hull in $E^2$ and it is easy to implement it as well. The experimental tests were made on a squared interval that is considered to be the worst case for testing of algorithm properties as far as the computational time is concerned. For oblong intervals, the proposed algorithm runs even faster. Another nice property is its extensibility to the $E^k$ case on the contrary to algorithms based on convex hull.

It is necessary to note that the presented algorithm is "output sensitive" type, so it is not convenient for an extremely special cases, when all points are points of a circle as $N (N-1)/2$ points have the same distance etc.. Nevertheless even in this case the algorithm is faster than the "Brute Force" and algorithms based on a convex hull construction. It can be seen that the presented algorithm can be extended for the $E^3$ data sets as well.

**ACKNOWLEDGMENTS**

The author would like to express his thanks to students of the University of West, especially to students Vit Ondracka and others for experimental verification of the proposed algorithm and an extensive testing. Thanks belong also to colleagues at the University of Ostrava, University of Jinan (China) and University of Hangzhou (China) for constructive discussions that helped to finish this work. Many thanks are given to anonymous reviewers for their valuable comments and suggestions that significantly contributed to the improvement of this paper.